\documentclass[12pt,preprint2]{emulateapj}

\shorttitle{Small-Scale Anisotropy of Ultrahigh Energy Cosmic Rays}
\shortauthors{The HiRes Collaboration}

\begin{document}

\title{Study of Small-Scale Anisotropy of Ultrahigh Energy Cosmic Rays
Observed in Stereo by HiRes}

\author{
R.U.~Abbasi,\altaffilmark{1}
T.~Abu-Zayyad,\altaffilmark{1}
J.F.~Amann,\altaffilmark{2}
G.~Archbold,\altaffilmark{1}
R.~Atkins,\altaffilmark{1}
J.A.~Bellido,\altaffilmark{3}
K.~Belov,\altaffilmark{1}
J.W.~Belz,\altaffilmark{4}
S.~BenZvi,\altaffilmark{5}
D.R.~Bergman,\altaffilmark{6}
J.H.~Boyer,\altaffilmark{5}
G.W.~Burt,\altaffilmark{1}
Z.~Cao,\altaffilmark{1}
R.W.~Clay,\altaffilmark{3}
B.M.~Connolly,\altaffilmark{5}
B.R.~Dawson,\altaffilmark{3}
W.~Deng,\altaffilmark{1}
Y.~Fedorova,\altaffilmark{1}
J.~Findlay,\altaffilmark{1}
C.B.~Finley,\altaffilmark{5}
W.F.~Hanlon,\altaffilmark{1}
C.M.~Hoffman,\altaffilmark{2}
M.H.~Holzscheiter,\altaffilmark{2}
G.A.~Hughes,\altaffilmark{6}
P.~H\"{u}ntemeyer,\altaffilmark{1}
C.C.H.~Jui,\altaffilmark{1}
K.~Kim,\altaffilmark{1}
M.A.~Kirn,\altaffilmark{4}
B.C.~Knapp,\altaffilmark{5}
E.C.~Loh,\altaffilmark{1}
M.M.~Maestas,\altaffilmark{1}
N.~Manago,\altaffilmark{7}
E.J.~Mannel,\altaffilmark{5}
L.J.~Marek,\altaffilmark{2}
K.~Martens,\altaffilmark{1}
J.A.J.~Matthews,\altaffilmark{8}
J.N.~Matthews,\altaffilmark{1}
A.~O'Neill,\altaffilmark{5}
C.A.~Painter,\altaffilmark{2}
L.~Perera,\altaffilmark{6}
K.~Reil,\altaffilmark{1}
R.~Riehle,\altaffilmark{1}
M.D.~Roberts,\altaffilmark{8}
M.~Sasaki,\altaffilmark{7}
S.R.~Schnetzer,\altaffilmark{6}
M.~Seman,\altaffilmark{5}
K.M.~Simpson,\altaffilmark{3}
G.~Sinnis,\altaffilmark{2}
J.D.~Smith,\altaffilmark{1}
R.~Snow,\altaffilmark{1}
P.~Sokolsky,\altaffilmark{1}
C.~Song,\altaffilmark{5}
R.W.~Springer,\altaffilmark{1}
B.T.~Stokes,\altaffilmark{1}
J.R.~Thomas,\altaffilmark{1}
S.B.~Thomas,\altaffilmark{1}
G.B.~Thomson,\altaffilmark{6}
D.~Tupa,\altaffilmark{2}
S.~Westerhoff,\altaffilmark{5}
L.R.~Wiencke,\altaffilmark{1}
A.~Zech\altaffilmark{6}\\
(The High Resolution Fly's Eye Collaboration)}


\altaffiltext{1}{University of Utah,
Department of Physics and High Energy Astrophysics Institute,
Salt Lake City, UT 84112, USA.}

\altaffiltext{2}{Los Alamos National Laboratory,
Los Alamos, NM 87545, USA.}

\altaffiltext{3}{University of Adelaide, Department of Physics,
Adelaide, SA 5005, Australia.}

\altaffiltext{4}{University of Montana, Department of Physics and Astronomy,
Missoula, MT 59812, USA.}

\altaffiltext{5}{Columbia University, Department of Physics and
Nevis Laboratories, New York, NY 10027, USA: finley@phys.columbia.edu,
westerhoff@nevis.columbia.edu.}

\altaffiltext{6}{Rutgers --- The State University of New Jersey,
Department of Physics and Astronomy, Piscataway, NJ 08854, USA.}

\altaffiltext{7}{University of Tokyo,
Institute for Cosmic Ray Research,
Kashiwa City, Chiba 277-8582, Japan.}

\altaffiltext{8}{University of New Mexico,
Department of Physics and Astronomy,
Albuquerque, NM 87131, USA.}

\begin{abstract}
The High Resolution Fly's Eye (HiRes) experiment is an air
fluorescence detector which, operating in stereo mode, 
has a typical angular resolution of $0.6^{\circ}$ and
is sensitive to cosmic rays with energies above $10^{18}$\,eV.  
HiRes is thus an excellent
instrument for the study of the arrival directions of ultrahigh energy
cosmic rays.  We present the results of a search for anisotropies 
in the distribution of arrival directions on small scales 
($<5^{\circ}$) and at the highest energies ($>10^{19}$~eV).  The search
is based on data recorded between 1999 December and 2004 January, with 
a total of 271 events above $10^{19}$\,eV.
No small-scale anisotropy is found, and the strongest clustering
found in the HiRes stereo data is consistent at the 52\,\% level with the
null hypothesis of isotropically distributed arrival directions.
\end{abstract}


\section{Introduction}

Identifying the sources of ultrahigh energy cosmic rays remains
one of the central challenges in astrophysics.  After three
decades of systematic searches for the origin of these particles,
source identification still remains elusive.
Sky maps of cosmic ray arrival directions at all energies 
are generally isotropic, with no obvious source or source region 
standing out.

A direct way to search for sources of ultrahigh energy cosmic rays
is to analyze the distribution of their arrival directions for 
small-scale clustering.
Any significant clustering in arrival directions could be
evidence of nearby, compact sources, whereas the lack of clustering is 
consistent with models in which ultrahigh energy cosmic ray sources 
are distributed at large distances from our Galaxy.

Arrival directions do not necessarily point 
back to sources, as charged cosmic ray primaries suffer deflections 
traveling through Galactic and intergalactic magnetic fields.  The 
strength and orientation of these fields is not well established, so 
the size and direction of the deflection is difficult to ascertain.  
However, since
the Larmor radius increases with energy, the possibility of observing 
small-scale anisotropy associated with cosmic rays pointing
back to their origins is expected to grow.

Indeed, small-scale clustering of cosmic ray arrival directions at the
highest energies has been previously claimed.  The AGASA 
(Akeno Giant Air Shower 
Array) experiment reported possible clustering in their sample of 
events with energies above $4\cdot10^{19}$\,eV \citep{agasa1996}. 
The analysis has been updated several times \citep{agasa1999, agasa2001, 
agasa2003}, most recently reporting six clusters (five 
doublets and one triplet) in a sample of 59 events, where a cluster is 
defined as a set of events with angular separation less than $2.5^{\circ}$.  
The chance probability of this signal was reported to be less than 
$10^{-4}$ \citep{agasa2003}.

Given the potential importance of this result for our understanding
of the origin of cosmic rays,
it is crucial to test the claim that clustering is a feature of
cosmic ray arrival directions with independent experimental data.
Since 1999, the High 
Resolution Fly's Eye (HiRes) air fluorescence experiment has been 
operating in stereo mode, collecting 
data of unprecedented quality on the arrival direction, 
energy, and composition of ultrahigh energy cosmic rays.
In this Letter, we report results 
of a search for small-scale anisotropy in the arrival directions of 
ultrahigh energy cosmic rays observed by the HiRes stereo detector 
between 1999 December and 2004 January.

\section{The HiRes Detector}

HiRes is an air fluorescence experiment with two sites
(HiRes\,1\,\&\,2) at the US Army Dugway Proving Ground in the
Utah desert ($112^{\circ}$\,W longitude, $40^{\circ}$\,N latitude,
vertical atmospheric depth $860\,{\mathrm{g}}/{\mathrm{cm}}^{2}$).
The two sites are separated by a distance of 12.6\,km.

Each of the two HiRes ``eyes'' comprises several telescope units
monitoring different parts of the night sky.  With 22 (42)
telescopes with 256 photomultiplier tubes each at the first (second)
site, the full detector covers about $360^{\circ}$ ($336^{\circ}$)
in azimuth and $3^{\circ}-16.5^{\circ}$ ($3^{\circ}-30^{\circ}$) in
elevation above horizon.  Each telescope consists of a mirror with an 
area of about $5\,\mathrm{m}^{2}$ area for light collection and a 
cluster of photomultiplier tubes in the focal plane.

A cosmic ray primary interacting in the upper atmosphere induces an
extensive air shower which the detectors observe as it develops through
the atmosphere.  
The photomultiplier tubes triggered by the shower define an
arc on the sky, and, together with the position of the detector, the
arc determines the so-called shower-detector plane.  When an air shower
is observed in stereo, the shower trajectory is in principle simply the
intersection of the two planes.  This method can be further improved by also
taking advantage of the timing information of the tubes, and in our
analysis the shower geometry is determined by a global $\chi^2$
minimization using both the timing and pointing information of all tubes.
From measurements of laser tracks and stars in the field of view
of the cameras we estimate that the systematic error in the arrival
direction determination is not larger than $0.2^{\circ}$, mainly caused
by uncertainties in the survey of mirror pointing directions.

Various aspects of the HiRes detector and the 
reconstruction procedures are described in \citet{nim2002, star2002, 
matthews2003}.

\section{The HiRes Data Set}

While a ground array detector can operate year-round, night and day,
air fluorescence detectors can only be operated on dark,
moonless nights with good atmospheric conditions.  This limits
the duty cycle to about $10\,\%$.  However, several years of observation
yield a data set with a relatively smooth distribution in
sidereal time, modulated by an overall seasonal variation in exposure.

\begin{figure*}
\epsscale{.8}
\plotone{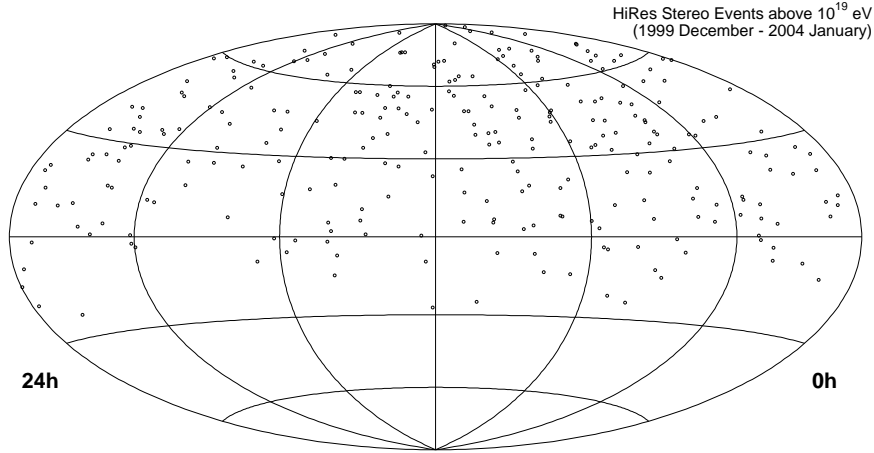}
\caption{Skymap (in equatorial coordinates) of the 271 
HiRes stereo events above
$10^{19}$\,eV examined in this study.  The typical
error radius of $0.6^{\circ}$ is used for all events.\label{skyplot}}
\end{figure*}

For the present analysis, we subject the HiRes stereo event sample 
to the following quality cuts.  We require a minimum track length 
of $3^{\circ}$ in each detector, an estimated angular uncertainty
in both azimuth and zenith angle of less than
$2^{\circ}$, and a zenith angle less than $70^{\circ}$.  We additionally 
require an estimated energy uncertainty
of less than $20\%$
and $\chi^2/{\rm dof}<5$ for both the energy and the geometry
fit.  Weather conditions which reduce the quality of the
data are cut implicitly in the above sample, 
rather than by explicit weather cuts. 
A total of 271 events above $10^{19}$\,eV 
pass the selection criteria.
A sky map in equatorial coordinates of the arrival directions of 
these events is shown in Figure\,\ref{skyplot}.

The angular resolution of HiRes is determined using simulated showers.
We use a full detector simulation 
of proton showers generated with CORSIKA 6 \citep{corsika1998} 
using QGSJET for the first interaction.  
Applying the same cuts to the simulation data which are 
applied to the real data, 68\,\% of
all showers generated at $10^{19}$\,eV are reconstructed within less than
$0.57^{\circ}$ of the true shower direction.  
The angular resolution depends weakly on energy, with the 68\,\% error
radius growing to $0.61^{\circ}$ and $0.69^{\circ}$ 
for showers generated at $4\cdot10^{19}$\,eV and $10^{20}$\,eV, respectively,
because at higher energy, showers are on average farther away.
The angular resolution is essentially constant in zenith and
azimuth angle of the arrival direction, varying by less than $0.1^{\circ}$.

\begin{figure}
\epsscale{.92}
\plotone{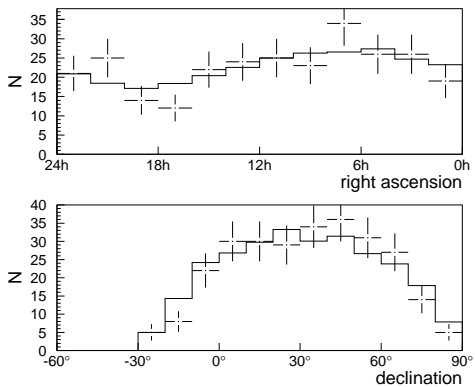}
\caption{
Right ascension and declination of events above $10^{19}$\,eV 
observed from 1999 December through 2004 January.  (Data---points with
error bars; Monte Carlo---solid line.)  For right ascension, 
$\chi^2/{\rm dof}=0.77$; for declination, $\chi^2/{\rm dof}=0.73$.
\label{ra_dec}}
\end{figure}

Using the same simulation described above, we generate an
isotropic distribution of showers with a differential spectral index
$\alpha=-3.0$ in energy, and use 
the resulting distribution of reconstructed Monte
Carlo events to determine the detector acceptance in zenith
and azimuth.
We then randomly match the local coordinates of these events
 with times during which the
detector was operating in order to generate an 
exposure map in equatorial coordinates.
Figure\,\ref{ra_dec} shows the distributions of the data and 
Monte Carlo events in right ascension and declination.

\section{Method}

We search for small-scale clustering by performing an auto-correlation
scan in energy and angular separation.  Essentially, we consider the
set of $N$ events above energy $E$, count the number
of pairs $n_p$ separated by less than $\theta$, and evaluate the probability 
$P(N,\theta)$ of finding this number or more pairs, given $N$ and $\theta$.  
We repeat this for a range of values for $E$ and $\theta$, and use
 the smallest probability $P_{min}$ found in the scan to identify
the strongest clustering signal.  We estimate the statistical significance 
$P_{ch}$ of this signal by performing identical scans over simulated sets of 
isotropically distributed data, counting the fraction of simulated
sets which yield the same or smaller value for $P_{min}$.

The virtue of this approach is that by letting the
energy threshold vary, we let the scan itself determine the optimal
balance between the better statistics of the low energy data set and 
the (presumably) smaller angular deflections at high energies.
Furthermore, we can simultaneously look for clustering
both at the angular scale identified by AGASA and at smaller scales that 
take advantage of the HiRes angular resolution.  The statistical penalty
for performing multiple searches is accounted for in the final
evaluation of the significance $P_{ch}$.

We note that, just as in the usual two-point correlation function, 
higher-order multiplets are counted by the individual number of 
pairs which they contain.

To determine the probabilities $P(N,\theta)$, we generate a large
number of simulated data sets (typically $10^7$) 
corresponding to an isotropic distribution of cosmic rays.
Specifically, we generate an event with a
random arrival direction in equatorial coordinates, and accept 
that event into the simulated data set 
with a probability proportional to the HiRes exposure in that region
of the sky.  We then construct a table of values $P_{MC}$, where
$P_{MC}(N,\theta,n)$ is the fraction of data sets in which the first
$N$ events contain exactly $n$ pairs separated by less than
$\theta$.  Then the probability $P(N,\theta)$ for observing $n_p$ 
or more pairs at $(N, \theta)$ is simply:
\begin{equation} 
   P(N, \theta) = \sum_{n=n_p}^{\infty} P_{MC}(N,\theta, n)
     = 1 - \sum_{n=0}^{n_p-1} P_{MC}(N,\theta, n). 
\end{equation}
For some combination $N_c$ and $\theta_c$, $P$ has a minimum:
$P_{min} = P(N_c, \theta_c)$.  We identify this as the strongest
potential clustering signal.  To determine the statistical significance,
we perform the same scan over $n_{MC}$ Monte Carlo data sets, 
finding the minimum
probability $P_{min}^{i}=P^{i}(N^i_c,\theta^i_c)$ for each trial
and counting the number of trials $n_{MC}^{\ast}$ for which
$P_{min}^{i}\le P_{min}$.  The significance is finally identified
as:
\begin{equation}
P_{ch} = \frac{n_{MC}^{\ast}}{n_{MC}},
\end{equation}
that is, the chance probability of observing 
the value $P_{min}$ or less in an isotropic distribution.

The scan is performed over the total set of $N=271$ events and 
over angular separations $\theta$ from $0^{\circ}$ to
$5^{\circ}$ in increments $\Delta \theta$ of $0.1^{\circ}$.
Rather than use an arbitrary fixed increment $\Delta E$ of energy,
we increase the energy threshold one event at a time ($\Delta N = 1$).
These search parameters were chosen {\it a priori}.  
While the results inevitably depend on the exact choices,
the dependence is relatively small \citep[see][for details and examples]
{finley2004}.

To demonstrate the effectiveness of this method and the sensitivity of the
HiRes detector, we apply this technique to simulated data with clusters.  
First, we generate
a set of 271 events with the HiRes exposure for isotropic arrival directions.
We then insert $m$ pairs of events among the $N_H$ highest energy events
in the set to simulate clustering above a specific energy threshold.

To create a pair, we pick a point in the sky for the source location and 
generate two events with arrival directions deviating from the source 
location according to a Gaussian distribution described below.  These 
artificial cluster positions are chosen at random, but their distribution 
is forced to reflect the overall exposure of the HiRes detector, so
that regions with higher exposure are more likely to contain a cluster.
The pair of events is then added to the original isotropic data set, replacing
two of the original events in the set.  This is repeated until
$m$ pairs have been inserted.  The set may contain more than $m$ pairs due
to chance.

For simplicity, we use a circular Gaussian distribution for the smearing
of arrival directions around the source location.  The width of the
distribution $\sigma_R$ can be set equal to the angular resolution of the
detector, or it can be set to a larger value to simulate additional smearing
by magnetic fields.  
(Note that for the Gaussian distribution 
$P(\theta)=(\theta/\sigma^2) e^{-\theta^2/ 2 \sigma^2}$, 
the value $\theta=1.515\,\sigma$ encloses $68\%$ of the distribution.
We therefore define $\sigma_R = 1.515\,\sigma$.)

Table\,\ref{simtable} shows the results of these simulations using
the detector resolution ($\sigma_R=0.6^{\circ}$), as well as three
times the detector resolution ($\sigma_R=1.8^{\circ}$) to simulate 
additional smearing by magnetic fields.
For each choice of $N_H$, $m$, and $\sigma_R$, we generate $10^4$ data
sets, and scan them with the procedure described above to find
a distribution of values for the significance $P_{ch}$.
The median and $90^{\rm th}$ percentile values of this 
distribution are indicated in Table\,\ref{simtable}.

\begin{deluxetable}{cccccc}
\tablecolumns{5}
\tablewidth{0pt}
\tablecaption{Results for Simulated Clusters\label{simtable}}
\tablehead{ & & \multicolumn{2}{c}{$\sigma_R=0.6^{\circ}$} &
                \multicolumn{2}{c}{$\sigma_R=1.8^{\circ}$} \\
  \colhead{$N_H$ \tablenotemark{a}} & \colhead{$m$} & 
  \colhead{median $P_{ch}$}  &  \colhead{90\% $P_{ch}$} &
  \colhead{median $P_{ch}$}  &  \colhead{90\% $P_{ch}$} }
\startdata
27  &  2  &  0.018               &  0.090  &     0.13  &  0.48 \\
    &  3  &  $2.5\cdot 10^{-3}$  &  0.013  &     0.050 &  0.25 \\
    &  4  &  $3.1\cdot 10^{-4}$  &  $1.5\cdot 10^{-3}$  &     0.016 &  0.11 \\

47  &  3  &  0.011               &  0.067  &     0.12  &  0.47 \\
    &  4  &  $1.9\cdot 10^{-3}$  &  0.012  &     0.059 &  0.32 \\
    &  5  &  $3.3\cdot 10^{-4}$  &  $2.2\cdot 10^{-3}$  &     0.029 &  0.18 \\

89  &  4  &  0.016               &  0.11   &     0.16  &  0.59 \\
    &  6  &  $1.0\cdot 10^{-3}$  &  0.012  &     0.071 &  0.38 \\
    &  8  &  $1.1\cdot 10^{-4}$  &  $7.3\cdot 10^{-4}$  &     0.025 &  0.20 \\
\enddata
\tablenotetext{a}{$N_H=$ 27, 47, and 89 events corresponds to simulated
clustering above energy thresholds 40 EeV, 28 EeV, and 20 EeV, respectively.}
\end{deluxetable}

The table shows, for example, that for a clustering signal on the  
$\sigma_R=0.6^{\circ}$ scale, even three pairs among
the 47 highest energy events would typically result in $P_{ch}=1.1\%$.
The table also shows that three such pairs would result in $P_{ch}<6.7\%$ for
90\% of the simulated sets.  Thus, an actual value of $P_{ch}>6.7\%$ could
be used to exclude the possibility that sources contributed three such pairs 
at more than the 90\% confidence level.

These results demonstrate the sensitivity to clustering 
on small angular scales.

\section{Results and Discussion}

We perform the scan on the HiRes stereo sample of
271 events above $10^{19}$\,eV.  Because
we start well below the $4\cdot 10^{19}$\,eV energy
associated with the AGASA clustering signal, our search should safely
encompass the energy region of interest even in the presence of a 
systematic energy shift of 30\% between the two experiments, 
as suggested by \citet{demarco2003}.
Starting at this energy
does not appreciably dilute the significance of a clustering
signal if one is found at higher energy, since the scan involves 
repeated searching with successively higher energy thresholds.
An additional motivation for starting at $10^{19}$\,eV 
is the fact that the HiRes angular resolution
($0.6^{\circ}$) is much sharper at this energy than AGASA's ($2.8^{\circ}$)
\citep{agasa1999}.

\begin{figure}
\epsscale{.95}
\plotone{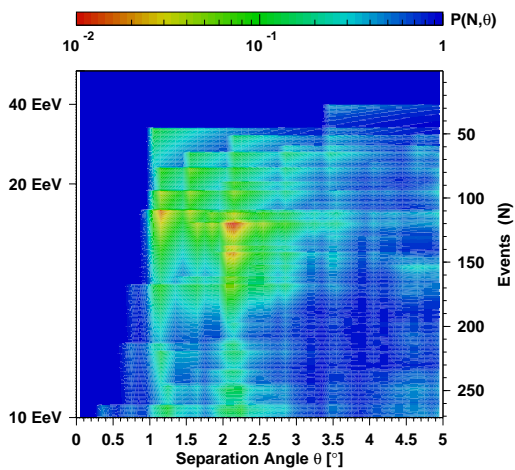}
\caption{Autocorrelation scan of the HiRes data set above
$10^{19}$\,eV.  $P(N,\theta)$ is the probability
of obtaining the same or greater number of pairs as is actually
observed in the data using a maximum separation angle $\theta$ and
searching among the $N$ highest-energy events.  These probabilities
do not include the statistical penalty due to scanning.
\label{scan}}
\end{figure}

The results of the 
scan are shown in Figure\,\ref{scan}.
The strongest clustering signal ($P_{min}=1.9\%$) 
is observed using the energy threshold
$E_c = 1.69\cdot 10^{19}$\,eV where we observe $n_p=10$ pairs 
separated by less than 
$\theta_c=2.2^{\circ}$ within a set of $N_c=120$ events.  
The statistical significance of this result corresponds to $P_{ch}=52\%$.

The HiRes stereo data above $10^{19}$\,eV is therefore 
consistent with the null hypothesis of isotropic arrival directions.

Comparison with the AGASA clustering result is not straightforward.
The HiRes stereo event sample above $4\cdot 10^{19}$\,eV is still smaller than 
AGASA's, though how much smaller depends critically on the level of agreement
in absolute energy scale for the two experiments.  
The possibility of a systematic energy shift of 30\% would imply that
above the rescaled energy threshold,
$(0.7)\cdot 4\cdot 10^{19}\,{\rm eV} = 2.8 \cdot 10^{19}\,{\rm eV}$,
HiRes has seen 47, rather than 27, events.
More importantly, there is the question of how many pairs an
independent data set might be expected to contain, given the lack of an
obvious source model and the widely varying estimates of the strength of
the AGASA clustering.  Without assuming a model and source strength, there
is no natural way to translate the AGASA observation of five doublets and
one triplet separated by less than $2.5^{\circ}$ into a meaningful
prediction for HiRes.

However, what can be tested using a statistically independent data set 
is the claim that significant
small-scale clustering is a general feature of ultrahigh energy cosmic ray
arrival directions.  The HiRes stereo data set does not support such a claim.  
We observe no
statistically significant evidence for clustering on any angular scale
up to $5^{\circ}$ at any energy threshold above $10^{19}$\,eV.

Comparing the observed value of $P_{ch}$ with the values obtained from
simulations in Section 4 (shown in Table\,\ref{simtable}), we note that if 
the current HiRes data above $4\cdot 10^{19}$\,eV contained two or more 
pairs of events contributed by compact sources at the angular resolution 
limit of the detector, then the
typical value of $P_{ch}$ would be 0.018 or less, and more than 90\% 
of the time
the value of $P_{ch}$ would be much smaller than the observed value of 0.52.

Results of searches for correlations with known astrophysical
source classes will be published in a separate paper.

\acknowledgments
The HiRes project is supported by the National Science Foundation under
contract numbers NSF-PHY-9321949, NSF-PHY-9322298, NSF-PHY-9974537,
NSF-PHY-0098826, NSF-PHY-0245428, by the Department
of Energy Grant FG03-92ER40732, and by the Australian Research Council.
The cooperation of Colonels E. Fisher and G. Harter, the US Army and
Dugway Proving Ground staff is appreciated.  
We thank the authors of CORSIKA for providing us with
the simulation code.

\end{document}